\documentstyle[12pt,moriond,epsf]{article} 
\def\HI{\hbox{H~$\scriptstyle\rm I\ $}}
\def\HII{\hbox{H~$\scriptstyle\rm II\ $}}
\def\nHI{{\rm HI}}
\def\nH{{\rm H}}
\def\nHII{{\rm HII}}

\def\HeII{\hbox{He~$\scriptstyle\rm II\ $}}

\def\HeIII{\hbox{He~$\scriptstyle\rm III\ $}}
\def\CIV{\hbox{C~$\scriptstyle\rm IV\ $}}

\def\kms{\,{\rm km\,s^{-1}}}
\def\kmsmpc{\,{\rm km\,s^{-1}\,Mpc^{-1}}}

\def\emunits{\,{\rm ergs\,s^{-1}\,Hz^{-1}\,Mpc^{-3}}}
\def\ndotunits{\,{\rm s^{-1}\,Mpc^{-3}}}
\def\ndotun{\,{\rm phot\,s^{-1}}}
\def\msun{\,{\rm M_\odot}}

\def\sfrd{\,{\rm M_\odot\,yr^{-1}\,Mpc^{-3}}}
\def\sfr{\,{\rm M_\odot\,yr^{-1}}}

\def\Lya{Ly$\alpha\ $}

\def\AB{{\rm AB}}
\def\spose#1{\hbox to 0pt{#1\hss}}
\def\lta{\mathrel{\spose{\lower 3pt\hbox{$\mathchar"218$}}
     \raise 2.0pt\hbox{$\mathchar"13C$}}}
\def\gta{\mathrel{\spose{\lower 3pt\hbox{$\mathchar"218$}}
     \raise 2.0pt\hbox{$\mathchar"13E$}}}

\begin{document}
\heading{Star Formation at High Redshift: A Population of Early Dwarfs?}            

\author{P. Madau} {Space Telescope Science Institute, Baltimore, USA.}

\begin{moriondabstract}
The history of the transition from a neutral intergalactic medium (IGM) to one
that is almost fully ionized can reveal the character of cosmological ionizing
sources and set important constraints on the stellar birthrate at high
redshifts. The hydrogen component in a highly inhomogeneous universe is
completely reionized when the number of photons emitted above 1 ryd in one
recombination time equals the mean number of hydrogen atoms.
If stellar sources are responsible for photoionizing the IGM at $z=5$, the 
rate of star formation at this epoch must be comparable or greater than the one 
inferred from optical observations of galaxies at $z\approx 3$, and the mean 
metallicity per baryon in the universe $\gta 1/500$ solar. In hierarchical
clustering scenarios, high-$z$ dwarfs (i.e. an early generation of stars in dark
matter halos with circular velocities $v_{\rm circ} \approx 50\,\kms$) are expected 
to be one of the main source of UV photons and heavy elements at early epochs. 
They would be very numerous, $\gta 0.2$ arcsec$^{-2}$, and faint, $I_{\rm AB}\gta
29.5\,$ mag: their detection may have to wait for the {\it Next Generation Space 
Telescope}. 
\end{moriondabstract}

\section{Introduction}

What keeps the universe ionized at $z=5$?
The existence of a filamentary, low-density intergalactic medium (IGM), which
contains the bulk of the hydrogen and helium in the universe, is predicted as a
product of primordial nucleosynthesis \cite{CST} and of hierarchical models of 
gravitational instability with ``cold dark matter'' (CDM) \cite{C94}, \cite{ZAN95},
\cite{H96}. The application of the Gunn-Peterson \cite{GP} constraint on the amount 
of smoothly distributed neutral material along the line of sight to distant objects 
requires the hydrogen component of the diffuse IGM to have been highly ionized 
by $z\approx 5$ \cite{SSG}, and the helium component by $z\approx 2.5$ \cite{DKZ}.
The plethora of discrete absorption systems which give origin to the \Lya forest in 
the spectra of background quasars are also inferred to be strongly photoionized. 
From QSO absorption studies we also know that neutral hydrogen accounts for only a 
small fraction, $\sim 10\%$, of the nucleosynthetic baryons at early epochs \cite{LWT}.
It thus appears that substantial sources of ultraviolet
photons were present at $z>5$, perhaps low-luminosity quasars \cite{HL98} or a first 
generation of stars in virialized dark matter halos with $T_{\rm vir}\sim 10^4-10^{5}
\,$K \cite{CR}, \cite{OG96}, \cite{HL97}, \cite{MR98}: early star
formation provides a possible explanation for the widespread existence of heavy
elements in the IGM \cite{C95}. More in general, establishing the character of
cosmological ionizing sources is an efficient way to constrain competing
models for structure formation in the universe, and to study the collapse and
cooling of small mass objects at early epochs. 

In this talk I will focus on the candidate sources of photoionization at 
early times and on the time-dependent reionization problem, i.e. on the history
of the transition from a neutral IGM to one that is almost fully ionized. Throughout 
this paper I will adopt an Einstein-de Sitter universe ($q_0=0.5$) with $H_0=
50h_{50}\, \kmsmpc$. The ideas described below have been developed in 
collaboration with F. Haardt and M. J. Rees \cite{MHR}. 

\section{Sources of Ionizing Radiation at High Redshifts}

\subsection{Quasars}

The existence of a decline in the space density of bright quasars at redshifts
beyond $\sim 3$ was first suggested by \cite{O82}, and has been since then
the subject of a long-standing debate. In recent years, several optical surveys
have consistently provided new evidence for a turnover in the QSO counts \cite{HS90},
\cite{WHO}, \cite{Sc95}, \cite{KDC}. 
The interpretation of the drop-off observed in optically selected samples is
equivocal, however, because of the possible bias introduced by dust obscuration
arising from intervening systems \cite{OH84}. Radio emission, on the 
other hand, is unaffected by dust, and it has recently been shown \cite{Sha} that 
the space density of radio-loud quasars also decreases strongly for $z>3$, 
demonstrating that the turnover is indeed real and that dust along the line of sight 
has a minimal effect on optically-selected QSOs (Figure 1). 

\begin{figure}
\epsfysize=10cm 
\hspace{3.5cm}\epsfbox{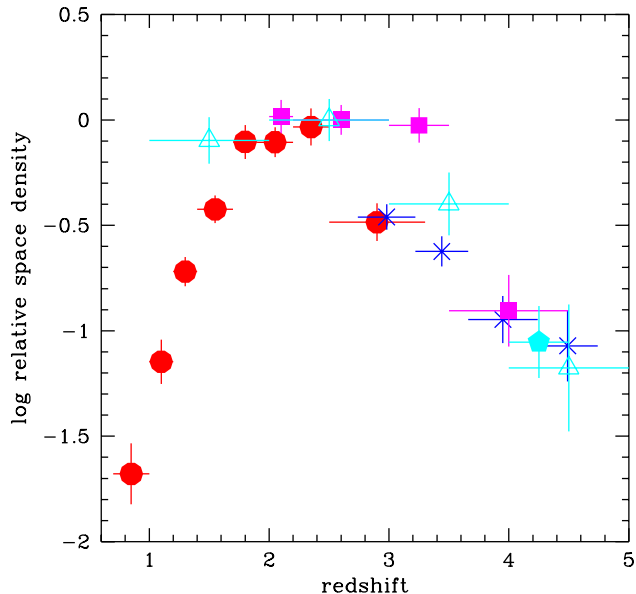}
\caption[h]{Comoving space density of bright QSOs as a function of redshift.
The data points with error bars are
taken from \cite{HS90} {\it (filled dots)}, \cite{WHO} {\it (filled 
squares)}, \cite{Sc95} {\it (crosses)}, and \cite{KDC} {\it (filled
pentagon)}. The points have been normalized to the $z=2.5$ space density of
quasars with $M_B<-26$ ($M_B<-27$ in the case of \cite{KDC}). The {\it empty
triangles} show the space density (normalized to the peak) of the Parkes
flat-spectrum radio-loud quasars with $P>7.2\times 10^{26}\,$ W Hz$^{-1}$
sr$^{-1}$ \cite{H98}. 
\label{fig1}}
\end{figure}
The QSO emission rate of hydrogen ionizing photons per unit comoving volume can
be written as \cite{M98}
\begin{equation}
\dot{\cal N}_Q(z)=(1.45\times 10^{50}\,\ndotunits) (1+z)^{\alpha_s-1}~{e^{\zeta
z} (1+e^{\xi z_*})\over e^{\xi z} + e^{\xi z_*}}{(1-4^{-\alpha_s})\over
\alpha_s} \left({4400\over 912}\right)^{-\alpha_s},   \label{eq:NdQ}
\end{equation}
where $\alpha_s$ is the slope of the quasar spectral energy
distribution, and the best-fit parameters are $z_*=1.9$, $\zeta=2.58$, and $\xi=3.16$. 
It is important to notice that the procedure adopted to derive this 
quantity implies a large correction for incompleteness at high-$z$. With a fit 
to the quasar luminosity function (LF)  which goes as $\phi(L)\propto 
L^{-1.64}$ at the faint end \cite{P95}, the
contribution to the emissivity converges rather slowly, as $L^{0.36}$. At
$z=4$, for example, the blue magnitude at the break of the LF is $M_*\approx 
-25.4$, comparable or slightly fainter than the limits of current high-$z$  
QSO surveys. A large fraction, about 90\% at $z=4$ and even higher at 
earlier epochs, of the ionizing emissivity in our model is therefore 
produced by quasars that have not been actually observed, and are
assumed to be present based on an extrapolation from lower redshifts. The 
value of $\dot{\cal N}_Q$ obtained by including the contribution from 
{\it observed} quasars only would be much smaller at high redshifts than 
shown in Figure 2. While it is also possible that an excess of low-luminosity 
QSOs, relative to the best-fit LF, could actually boost the estimated ionizing 
emissivity at early epochs, the observed lack of red, unresolved faint objects 
in the {\it Hubble Deep Field} seems to argue against models where the quasar 
LF steepens significantly with lookback time \cite{HML}.

\subsection{Star-forming Galaxies}

Galaxies with ongoing star-formation are another obvious source of Lyman
continuum photons. The tremendous progress in our understanding of faint galaxy
data made possible by the recent identification of star-forming galaxies at
$2\lta z\lta 4$ in ground-based surveys \cite{S96a} and in the {\it Hubble Deep
Field} (HDF) \cite{S96b}, \cite{M96}, \cite{Low97} has provided new
clues to the long-standing issue of whether galaxies at high redshifts can 
provide a significant contribution to the ionizing background flux. Since the 
rest-frame UV continuum at 1500 \AA\ (redshifted into the visible band for a
source at $z\approx 3$) is dominated by the same short-lived, massive stars
which are responsible for the emission of photons shortward of the Lyman edge,
the needed conversion factor, about one ionizing photon every 10 photons at
1500 \AA, is fairly insensitive to the assumed IMF and is independent of the
galaxy history for $t\gg 10^{7.3}\,$ yr. 
A composite ultraviolet luminosity function of Lyman-break galaxies at 
$z\approx 3$ has been recently derived by \cite{D98}. It is based
on spectroscopically (about 375 objects) and photometrically selected 
galaxies from the ground-based and HDF samples, and spans about a factor
50 in luminosity from the faint to the bright end. Because of the uncertanties
that still remain in the rescaling of the HDF data points to the ground-based 
data, the comoving luminosity density at 1500 \AA\ is estimated to vary 
within the range 1.6 to $3.5\times 10^{26}\emunits$. The ``best guess'' 
Schechter fit gives $\log\ (\phi_*/{\rm Gpc^{-3}})=6.1$, $\alpha=1.38$, and
$M_\AB^*(1500)=-20.95$ \cite{D98}, the magnitude at the ``break''
corresponding to a star-formation rate slightly in excess of $10\sfr$ (Salpeter IMF). 

Figure 3 shows the Lyman continuum luminosity function of galaxies at $z\approx 3$ 
(at all ages $\gta 0.1$ Gyr one has $L(1500)/L(912)\approx 6$ for a Salpeter
mass function and constant star formation rate \cite{BC98}), 
compared to the distribution of QSO luminosities at the same redshift. The 
comoving ionizing emissivity due to Lyman-break galaxies is $4.2\pm 1.5
\times 10^{25}\emunits$, between 2 and 4 times higher than the 
estimated quasar contribution at $z=3$. 

This number neglects any correction for dust extinction and intrinsic
\HI absorption. While it has been pointed out by many authors \cite{Me97},
\cite{Pe98}, \cite{D98} that the colors of Lyman-break galaxies are redder than
expected in the case of dust-free star-forming objects, the prescription for a
``correct'' de-reddening is still unknown at present (note that redder spectra
may also results from an aging population or an IMF which is rather poor in
massive stars). A Salpeter IMF, $E_{\rm B-V}=0.1$ model with SMC-type dust in a
foreground screen, for example, has been found to reproduce quite well the
rest-frame ultraviolet colors of the HDF ``UV dropouts'' \cite{M98}. In 
this model the color excess $E_{912-1500}= 1.64E_{\rm B-V}$ is rather small
and can be safely neglected in correcting from observed rest-frame far-UV to
the Lyman edge. For typical dust-to-gas ratios, however, it is the \HI
associated with dust that would greatly reduce the flux of Lyman continuum 
photons able to escape into the intergalactic space. The data points plotted 
in Figure 2 assume a value of $f_{\rm esc}=0.5$ for the unknown fraction of 
ionizing photons which escapes the galaxy \HI layers into the intergalactic 
medium \cite{MS96}. The possible existence
of a numerous population of galaxies below the detection threshold, i.e. having
star formation rates $<0.5\sfr$, with a space density well in excess of
that predicted by extrapolating to faint magnitudes the $\alpha=1.38$ best-fit
Schechter function, will be discussed below.
\begin{figure}
\epsfysize=10cm 
\hspace{3.cm}\epsfbox{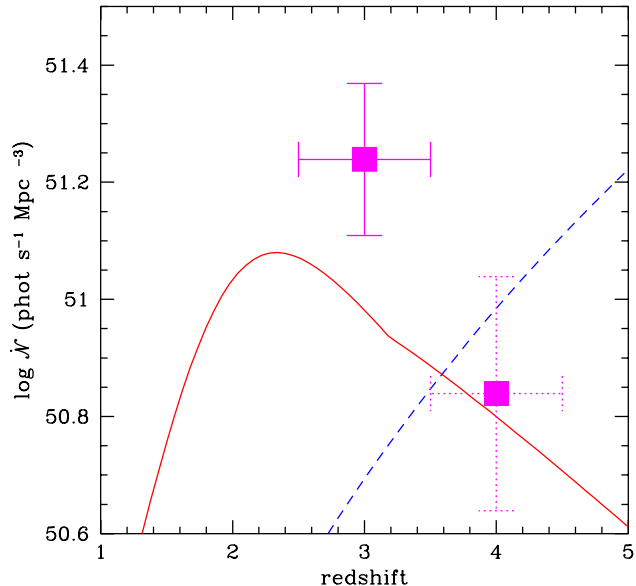}
\caption[h]{Comoving emission rate of hydrogen Lyman continuum photons ({\it solid
line}) from QSOs, compared with the minimum rate ({\it dashed line}) which is
needed to fully ionize a fast recombining (with clumping factor $C=30$) 
Einstein--de Sitter universe with $h_{50}=1$ and $\Omega_b=0.08$. Models based on 
photoionization by quasar sources appear to fall short at $z=5$. See \cite{MHR}
for details on the assumed
quasar luminosity function and spectral energy distribution.  The data points
with error bars show the estimated contribution of star-forming galaxies 
at $z\approx 3$ and, with significantly larger uncertainties, at $z\approx 4$. 
The fraction 
of Lyman continuum photons which escapes the galaxy \HI layers into the 
intergalactic medium is taken to be $f_{\rm esc}=0.5$.
\label{fig2}} 
\end{figure}
The LF of Lyman-break galaxies at $z\gta 4$ is highly uncertain. An analysis
of the $B$-band dropouts in the HDF -- candidate star-forming objects at
$3.5<z<4.5$ -- seems to imply a decrease in the comoving UV galaxy emissivity
by about a factor of 2.5 in the interval $2.75\lta z\lta 4$ \cite{M96},
\cite{M98}. In this sense star-forming galaxies with SFR in excess of 
$0.5\sfr$ may 
have a negative evolution with lookback time similar to the one observed in 
bright QSOs, but the error bars are still rather large. Adopting a $L(1500)$ 
to $L(912)$ conversion factor of 6, we estimate a comoving ionizing emissivity 
of $1.7\pm 1.1 \times 10^{25} f_{\rm esc} \emunits$ at $z\approx 4$. One 
should note that, while a population of highly reddened galaxies at high 
redshifts would be missed by the dropout color technique (which isolates 
sources that have blue colors in the optical and a sharp drop in the 
rest-frame UV), it seems unlikely that very dusty objects (with $f_{\rm esc}
\ll 1$) would contribute in any significant manner to the ionizing 
metagalactic flux. 

\section{Reionization of the Universe}

In inhomogeneous reionization scenarios, the history of the transition from a
neutral IGM to one that is almost fully ionized can be statistically 
described by
the evolution with redshift of the {\it volume filling factor} or porosity
$Q(z)$ of \HII, \HeII, and \HeIII regions. The radiation emitted by spatially
clustered stellar-like and quasar-like sources -- the number densities and 
luminosities of which may change rapidly as a function of redshift -- 
coupled with
absorption processes in a medium with a time-varying clumping factor, all
determine the complex topology of neutral and ionized zones in the universe.
When $Q<<1$ and the radiation sources are randomly distributed, the ionized
regions are spatially isolated, every UV photon is absorbed somewhere in the
IGM, and the ionization process cannot be described as due to a statistically
homogeneous radiation field \cite{AW}. As $Q$ grows, the crossing of 
ionization fronts becomes more and more common, and the neutral phase shrinks 
in size until the reionization process is completed at the ``overlap'' epoch, 
when every point in space is exposed to Lyman continuum radiation and $Q=1$. 
\begin{figure}
\epsfysize=10cm 
\hspace{3.cm}\epsfbox{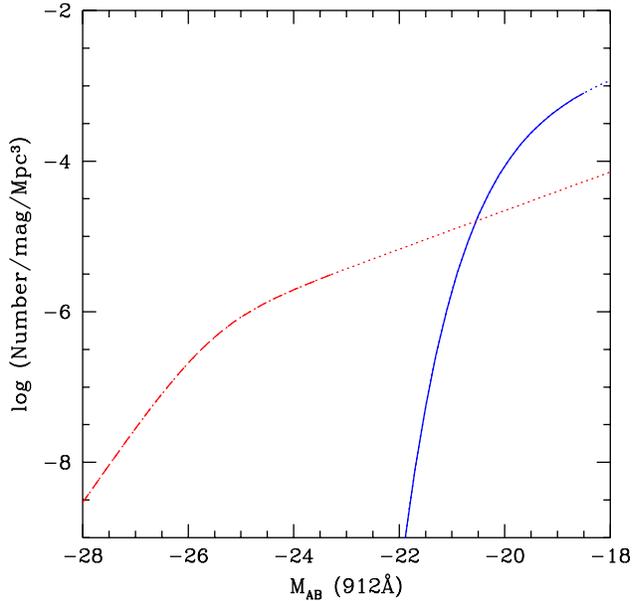}
\caption{The $912\,$\AA\ luminosity function of galaxies at $z\approx 3$ ({\it
solid line}), compared to the distribution of QSO luminosities at the same
redshift ({\it dashed line}). The latter has been derived assuming a spectral
slope of $\alpha_s=0.5$. The former assumes a Salpeter IMF with constant 
constant star formation rate (age=1 Gyr): $M_\AB(912\,$\AA)$=-19$ corresponds 
to a rate of $13\sfr$. The solid and dashed lines represent functional 
fits to the data points, and the dotted lines their extrapolation. 
\label{fig3}} 
\end{figure}

When an isolated point source of ionizing radiation turns on, the ionized
volume initially grows in size at a rate fixed by the emission of UV photons,
and an ionization front separating the \HII and \HI regions propagates
into the neutral gas. Most photons travel freely in the ionized bubble, and are
absorbed in a transition layer. The evolution of an expanding \HII region is 
governed by the equation 
\begin{equation}
{dV_I\over dt}-3HV_I={\dot N_{\rm ion}\over \bar{n}_\nH}-{V_I\over 
\bar{t}_{\rm rec}}, \label{eq:dVdt} 
\end{equation}
\cite{SG}, where $V_I$ is the proper volume of the ionized zone, $\dot N_{\rm ion}$ is
the number of ionizing photons emitted by the central source per unit time, $H$
is the Hubble constant, $\bar{n}_\nH$ is the mean hydrogen density of the expanding 
IGM, $\bar{n}_\nH(0)=1.7\times 10^{-7}$ $(\Omega_b h_{50}^2/0.08)$, and
\begin{equation}
\bar{t}_{\rm rec}=[(1+2\chi) \bar{n}_\nH \alpha_B\,C]^{-1}=0.3\, {\rm Gyr} 
\left({\Omega_b h_{50}^2 \over 0.08}\right)^{-1}\left({1+z\over 4}\right)^{-3} 
C_{30}^{-1} 
\end{equation}
is the volume-averaged gas recombination timescale. Here $\alpha_B$ is the 
recombination coefficient to the excited states of hydrogen,
$\chi$ the helium to hydrogen cosmic abundance ratio, $C\equiv \langle
n_\nHII^2\rangle/\bar{n}_\nHII^2$ is the ionized hydrogen clumping factor,
\footnote{This may be somewhat lower than the total gas clumping factor if 
higher density regions are less ionized \cite{GO97}.}\ and 
a gas temperature of $10^4\,$K has been assumed. Clumps which are dense
and thick enough to be self-shielded from UV radiation will stay neutral and
will not contribute to the recombination rate. An empirical determination of 
the  clumpiness of the IGM at high redshifts is hampered by our poor 
knowledge of the ionizing background intensity and the typical size and 
geometry of the absorbers. Numerical N-body/hydrodynamics simulations of 
structure formation in the IGM within the framework of CDM dominated 
cosmologies have 
recently provided a definite picture for the origin of intervening absorption 
systems, one of an interconnected network of sheets and filaments, with 
virialized systems located at their points of intersection.  In the simulations
of \cite{GO97}, for example, the clumping factor rises above 
unity when the collapsed fraction of baryons becomes non negligible, i.e. $z\lta 
20$, and grows to $C\gta 10$ (40) at $z\approx 8$ (5) (because of finite 
resolution effects, numerical simulations will actually underestimate 
clumping): the recombination timescale is much shorter than that for a uniform 
IGM, and always shorter than the expansion time.

When $\bar{t}_{\rm rec}\ll t$, the growth of the \HII region is
slowed down by recombinations in the highly inhomogeneous IGM, and its evolution
can be decoupled from the expansion of the universe. Just like in the static
case, the ionized bubble will fill its time-varying Str\"omgren sphere
after a few recombination timescales,
\begin{equation}
V_I={\dot N_{\rm ion}\bar{t}_{\rm rec}\over \bar{n}_\nH} 
(1-e^{-t/\bar{t}_{\rm rec}}). \label{eq:V} 
\end{equation}
One should point out that the use of a volume-averaged clumping factor in the 
recombination timescale is 
only justified when the size of the \HII region is large compared to the 
scale of the clumping, so that the effect of many clumps (filaments) within 
the ionized volume can be averaged over. This will be a good approximation 
either at late epochs, when the IGM is less dense and \HII zones have had time 
to grow, or at earlier epochs if the ionized bubbles 
are produced by very luminous sources like quasars or the stars within halos 
collapsing from very high-$\sigma$ peaks. The mean free path between absorbers having 
neutral columns 
$>N_\nHI$ is $0.8\,$ Mpc h$_{50}^{-1}\,[(1+z)/6)]^{-4.5}\,(N_\nHI/10^{15}\,{\rm 
cm^{-2}})^{0.5}$: it is only on scales greater than this value that the
clumping can be averaged over. On smaller scales underdense regions are 
ionized first, and only afterwards the UV photons start to gradually 
penetrate into the higher density gas.
 
With these caveats in mind, equation (\ref{eq:dVdt}) approximately holds for 
every isolated source of ionizing photons in the IGM. The filling factor of 
\HII regions in the 
universe, $Q_\nHII$, is then equal at any given instant $t$ to the integral
over cosmic time of the rate of ionizing photons emitted per hydrogen atom and
unit cosmological volume by all radiation sources present at earlier epochs, 
$\int_0^t \dot n_{\rm ion}(t')dt'/\bar{n}_\nH(t')$, {\it minus} the rate of 
radiative recombinations, $\int_0^t Q_\nHII(t')dt'/\bar{t}_{\rm rec}(t')$. 
Differentiating one gets
\begin{equation}
{dQ_\nHII\over dt}={\dot n_{\rm ion}\over \bar{n}_\nH}-{Q_\nHII\over 
\bar{t}_{\rm rec}}.  \label{eq:qdot}
\end{equation}
It is this simple differential equation -- and its equivalent for
expanding helium zones -- that statistically describes the transition
from a neutral universe to a fully ionized one, independently, for a given
emissivity, of the complex and possibly short-lived emission histories of
individual radiation sources, e.g., on whether their comoving space density is
constant or actually varies with cosmic time. In the case of a time-independent
clumping factor, equation (\ref{eq:qdot}) has formal solution 
\begin{equation}
Q_\nHII(t)=\int_0^t dt'\, {\dot n_{\rm ion}\over \bar{n}_\nH}\,
\exp\left(-{t'\over \bar{t}_{\rm rec}}+{t'^2\over \bar{t}_{\rm rec}t }\right),
\end{equation}
with $\bar{t}_{\rm rec}\propto t'^2$. At high 
redshifts, and for an IGM with $C\gg 1$, one can expand around $t$ to find 
\begin{equation}
Q_\nHII(t)\approx {\dot n_{\rm ion}\over \bar{n}_\nH}\bar{t}_{\rm rec}. 
\label{eq:qa}
\end{equation}
The porosity of ionized bubbles is then approximately given by the number
of ionizing photons emitted per hydrogen atom in one recombination time. In 
other words, because of hydrogen recombinations, only a fraction $\bar{t}_{\rm 
rec}/t$
($\sim$ a few per cent at $z=5$) of the photons emitted above 1 ryd is
actually used to ionize new IGM material. The universe is completely reionized
when $Q=1$, i.e. when 
\begin{equation}
\dot n_{\rm ion} \bar{t}_{\rm rec}=\bar{n}_\nH. 
\label{eq:qone}
\end{equation}
While this last expression has been derived assuming a constant
comoving ionizing emissivity and a time-independent clumping factor, it is also
valid in the case $\dot n_{\rm ion}$ and $C$ do not vary rapidly over a
timescale $\bar{t}_{\rm rec}$. 

\section{Dwarfs at High Redshift?}

As $\bar{t}_{\rm rec} \ll t$ at high redshifts, it is possible to compute 
at any given epoch a critical value for the photon emission rate per unit 
cosmological comoving volume, $\dot {\cal N}_{\rm ion}$,
independently of the (unknown) previous emission history of the universe: only
rates above  this value will provide enough UV photons to ionize the IGM by 
that epoch. One can then compare our determinations of $\dot {\cal N}_{\rm
ion}$ to the estimated contribution from QSOs and star-forming galaxies. 
Equation (\ref{eq:qone}) can then be rewritten as
\begin{equation}
\dot {\cal N}_{\rm ion}(z)={\bar{n}_\nH(0)\over \bar{t}_{\rm rec}(z)}=(10^{51.2}\,
\ndotunits)\, C_{30} \left({1+z\over 6}\right)^{3}\left({\Omega_b 
h_{50}^2\over 0.08}\right)^2. 
\label{eq:caln}
\end{equation}
The uncertainty on this critical rate is difficult to estimate, as it depends 
on the clumping factor of the IGM (scaled in the expression above to the 
value inferred at $z=5$ from numerical simulations \cite{GO97})
and the nucleosynthesis constrained baryon density. A quick exploration of the 
available parameter space indicates that the uncertainty on $\dot 
{\cal N}_{\rm ion}$ could easily be of order $\pm 0.2$ in the log. The 
evolution of the critical rate as a function of redshift is plotted in Figure 
2. While $\dot {\cal N}_{\rm ion}$ is comparable to the quasar contribution at 
$z\gta 3$, there is some indication of a significant deficit of Lyman 
continuum photons at $z=5$. For bright, massive galaxies to produce enough UV 
radiation at
$z=5$, their space density would have to be comparable to the one observed at
$z\approx 3$, with most ionizing photons being able to escape freely from the
regions of star formation into the IGM. This scenario may be in conflict with  
direct observations of local starbursts below
the Lyman limit showing that at most a few percent of the stellar ionizing
radiation produced by these luminous sources actually escapes into the IGM 
\cite{Le95}.\footnote{Note that, at $z=3$, Lyman-break galaxies would radiate 
more ionizing photons than QSOs for $f_{\rm esc}\gta 30\%$.}~If, on the other 
hand, faint QSOs with (say) $M_\AB=-19$ at rest-frame ultraviolet frequencies 
were to provide {\it all} the required ionizing flux, their 
comoving space density would be such ($0.0015\,$Mpc$^{-3}$) that about 50 
of them would expected in the HDF down to $I_\AB=27.2$.  At $z\gta 5$, they 
would appear very red in $V-I$ as the \Lya forest is shifted into the visible.
This simple model can be ruled out, however, as there is only a handful (7) 
of sources in the HDF with $(V-I)_\AB>1.5$ mag down to this magnitude limit.
  
It is interesting to convert the derived value of $\dot {\cal N}_{\rm ion}$  
into a ``minimum'' star formation rate per unit (comoving) volume, $\dot 
\rho_*$ (hereafter we assume $\Omega_bh_{50}^2=0.08$ and $C=30$):
\begin{equation}
{\dot \rho_*}(z)=\dot {\cal N}_{\rm ion}(z) \times 10^{-53.1} f_{\rm esc}^{-1}
\approx 0.013 f_{\rm esc}^{-1} \left({1+z\over 6}\right)^3\ \sfrd. \label{eq:sfr} 
\end{equation}  
The conversion factor assumes a Salpeter IMF with solar metallicity \cite{BC98}.
It can be understood by noting that, for each 1 $M_\odot$ of stars formed, 
8\% goes into massive stars with $M>20 M_\odot$ that dominate the 
Lyman continuum luminosity of a stellar population. At the end of the C-burning
phase, roughly half of the initial mass is converted into helium and carbon,
with a mass fraction released as radiation of 0.007. About 25\% of the energy
radiated away goes
into ionizing photons of mean energy 20 eV. For each 1 $M_\odot$ of stars
formed every year, we then expect 
\begin{equation} 
{0.08\times 0.5 \times 0.007 \times 0.25\times M_\odot c^2\over 
20 {\,\rm eV\,}} {1\over  {\rm 1\, yr}} \sim 10^{53}\ndotun
\end{equation} 
to be emitted shortward of 1 ryd. Note that the star formation density given in
equation (\ref{eq:sfr}) is comparable with the value directly ``observed'' 
(i.e., uncorrected for dust reddening) at $z\approx 3$ \cite{M98}, \cite{D98}.

The same massive stars that dominate the
Lyman continuum flux also manufacture and return most of the metals to the 
ISM. In the approximation of instantaneous recycling, the rate of ejection of 
newly sinthesized heavy elements which is required
to keep the universe ionized at redshift $z$ is, from equation (\ref{eq:sfr}),
\begin{equation}
{\dot \rho_Z}(z)=y(1-R){\dot \rho_*}(z)\gta 3.5\times 10^{-4}
\left({y\over 2 Z_\odot}\right) \left({1+z\over 6}\right)^3 f_{\rm esc}^{-1}\, 
\sfrd, \label{eq:mfr}
\end{equation}
where $y$ is the net, IMF-averaged ``yield'' of returned metals, $Z_\odot=0.02$,
and $R\approx 0.3$ is the mass fraction of a generation of stars that is 
returned to the interstellar medium. At $z=5$, and over a timescale of $\Delta 
t=0.5\,$ Gyr corresponding to a formation redshift $z_f=10$, such a rate would 
generate a mean metallicity per baryon in 
the universe of
\begin{equation}
\langle Z \rangle\approx {8\pi G {\dot \rho_Z}(5) \Delta t\over 3 H_0^2 
\Omega_b}\gta 0.002 \left({y\over 2 Z_\odot}\right) f_{\rm esc}^{-1}Z_\odot,
\end{equation}
comparable with the level of enrichment observed in the \Lya forest at
$z\approx 3$ \cite{So97}: more than 2\% of the present-day stars would need 
to have formed by $z\sim 5$. It has been recently suggested \cite{MR98} that a large 
number of low-mass galactic halos, expected to form at early times in hierarchical 
clustering models, might be responsible for photoionizing the IGM at these epochs. 
According to spherically-symmetric simulations \cite{TW96}, photoionization heating 
by the UV background flux 
that builds up after the overlapping epoch completely suppresses the cooling 
and collapse of gas inside the shallow potential wells of halos 
with circular velocities $\lta 35\,\kms$. Halos with circular speed 
$v_{\rm circ}=50\,\kms$, corresponding in top-hat spherical collapse to a 
virial temperature $T_{\rm vir}=0.5\mu m_p v_{\rm circ}^2/k\approx 10^{5}\,$K 
and halo mass $M=0.1v_{\rm circ}^3/GH\approx 4\times 10^9 [(1+z)/6]^{-3/2}
h_{50}^{-1}\, \msun$, appear instead largely immune to this external feedback
(but see \cite{NS97}). In these systems rapid cooling by 
atomic hydrogen can then take place and a significant fraction, 
$f\Omega_b$, of their total mass may be converted into stars over a timescale 
$\Delta t$ comparable to the Hubble time (with $f$ close 
to 1 if the efficiency of forming stars is high). If high-$z$ dwarfs with 
star formation rates $f\Omega_bM/\Delta t\sim 0.3 f_{0.5} \Delta t_{0.5}^{-1}\, 
\sfr$  were actually responsible for keeping the universe ionized at $z\sim 
5$, their comoving space density would have to be 
\begin{equation}
{0.013\, f_{\rm esc}^{-1}\sfrd\over 0.6 f {\, \rm M_\odot yr^{-1}}}
\sim 0.1 \left({f_{\rm esc} f\over 0.25}\right)^{-1} {\,\rm Mpc^{-3}},
\end{equation}
two hundred times larger than the space density of present-day galaxies 
brighter than $L^*(4400)$, and about five hundred times larger than that of 
Lyman-break objects at $z\approx 3$ with $M<M^*_\AB(1500)$,
i.e. with star formation rates in excess of $10\,\sfr$. Only a rather
steep luminosity function, with Schechter slope $\alpha\sim 2$, would
be consistent with such a large space density of faint dwarfs and, at the same
time, with the paucity of brighter $B$- and $V$-band dropouts observed in the
HDF. The number density on the sky would be $\approx 0.2\,$ arcsec$^{-2}$, 
corresponding to more than three thousands sources in the HDF. With a typical 
apparent magnitude at $z=5$ of $I_\AB \sim 29.5$ mag 
(assuming $f=0.5$), these might be too faint to 
be detected by the {\it HST}, but within the range of the proposed {\it Next
Generation Space Telescope} \cite{St98}.

These estimates are in reasonable agreement (although towards the low side) 
with the calculations of \cite{MR98}, who used the observed
metallicity of the \Lya forest as the starting point of their investigation. 
A higher density of sources -- which would therefore have to originate from 
lower amplitude peaks -- would be required if the typical efficiency of star 
formation and/or the escape fraction of ionizing photons were low, $(f, f_{\rm 
esc})\ll 1$. In this case the dwarfs could still be detectable  
if a small fraction of the gas turned into stars in very short bursts
(there would then be an extra parameter associated with their duty cycle, 
in addition to $f_{\rm esc}$ and $f$). A reduction of 
the star formation rate in halos with low circular velocities (necessary 
in hierarchical cosmogonies to prevent too many baryons from turning into stars
as soon as the first levels of the hierarchy collapse \cite{WF91})
may result from the heating and possible expulsion of the gas due to 
repeated supernova (SN) explosions after only a very small number of stars 
have formed. Recent numerical simulations \cite{Mac98} show, 
however, that metal-enriched material from SN ejecta is accelerated to 
velocities larger than the escape speed from such systems far more easily than 
the ambient gas. If the same population of dwarf galaxies that may keep the 
universe ionized at $z\approx 5$ were also responsible for polluting 
the IGM with heavy elements, it is interesting to ask whether these atoms 
could diffuse uniformly enough to account for the observations of weak but
measurable \CIV absorption lines associated with the \Lya forest clouds 
\cite{Lu98}, \cite{So97}. Our fiducial estimate of 0.1 sources 
per comoving Mpc$^3$ implies that, in order to pollute the entire IGM, the 
metal-enriched material would have to be expelled to typical distances of order
1 Mpc, which would require an ejection speed of about $200[(1+z)/6]^{1/2}\,
\kms$. In fact, the heavy elements may be restricted to filaments within only 
(say) 20 proper kpc (100-150 comoving kpc) of the halos, and in this case 
it would be enough to accelerate the outflowing metals to velocities comparable 
to the escape velocity, rather than to the higher speeds associated with 
SN ejecta. The required diffusion of heavy elements would be a more serious  
constraint if the relevant galaxies were rarer and more luminous, as would 
happen if they originated from 3-$\sigma$ peaks and star formation was 
postulated to occur with higher efficiency than in more typical peaks.

\vfill
\end{document}